\documentclass[useAMS,usenatbib]{mn2e}
\input journals.def
\usepackage{graphicx}
\usepackage[caption=false]{subfig}
\usepackage{mathtools}
\usepackage{times}
\usepackage[usenames,dvipsnames,svgnames,table]{xcolor}
\usepackage[T1]{fontenc}
\usepackage{aecompl}

\newcommand{\appropto}{\mathrel{\vcenter{
  \offinterlineskip\halign{\hfil$##$\cr
    \propto\cr\noalign{\kern2pt}\sim\cr\noalign{\kern-2pt}}}}}
\newcommand{\fov}{\mbox{Fo\hspace{-0.0833em}V } }
\newcommand{\fovmath}{\textrm{Fo\hspace{-0.0833em}V}}

\title[Subvirial Systems]{On the Effects of Subvirial Initial Conditions and the Birth Temperature of R136} 
\author[Caputo et al.]{Daniel~P.~Caputo$^1$\thanks{caputo@strw.leidenuniv.nl}, Nathan~de~Vries$^1$ and Simon~Portegies~Zwart$^1$\\
$^1$Leiden Observatory, Leiden University, PO Box 9513, 2300 RA Leiden, the Netherlands}

\begin{document}

\maketitle
\begin{abstract}
We investigate the effect of different initial virial temperatures, $Q$, on the dynamics of star clusters.  
We find that the virial temperature has a strong effect on many aspects of the resulting system, including among others: the fraction of bodies escaping from the system, the depth of the collapse of the system, and the strength of the mass segregation.
These differences deem the practice of using ``cold'' initial conditions no longer a simple choice of convenience.
The choice of initial virial temperature must be carefully considered as its impact on the remainder of the simulation can be profound.
We discuss the pitfalls and aim to describe the general behavior of the collapse and the resultant system as a function of the virial temperature so that a well reasoned choice of initial virial temperature can be made.
We make a correction to the previous theoretical estimate for the minimum radius, $R_{min}$, of the cluster at the deepest moment of collapse to include a $Q$ dependency, $R_{min}\approx Q + N^{(-1/3)}$, where $N$ is the number of particles.

We use our numerical results to infer more about the initial conditions of the young cluster R136.
Based on our analysis, we find that R136 was likely formed with a rather cool, but not cold, initial virial temperature ($Q\approx 0.13$).
Using the same analysis method, we examined 15 other young clusters and found the most common initial virial temperature to be between 0.18 and 0.25.
\end{abstract}

\begin{keywords}
galaxies:star clusters: general-galaxies: star clusters:individual:R136
\end{keywords}

\section{Introduction}
Subvirial systems are often used as initial conditions in numerical simulations for both physical and practical reasons.
Before the phase of gas expulsion, young stellar clusters must be formed subvirial, since the parent molecular cloud was roughly in virial equilibrium and supported by both gas pressure and (turbulent and systematic) velocities. 
The resultant stellar cluster is no longer supported by gas pressure, but only by the velocities of the stars, and therefore the energy balance must shift to subvirial.

In practice, subvirial conditions are also used to reduce the computational cost of reaching a mass segregated or otherwise relaxed system.
This is because with cold initial conditions mass segregation is established on a free-fall time-scale, but virial systems relax and reach mass segregation on a much longer time-scale.
Until now the consequence of changing the initial virial temperature has often been considered insignificant and so physical justification is not given.

If for example, an experiment is designed to investigate mergers \citep{SPZ1999,Bedorf2013} (or another physical phenomenon preferentially occurring in mass segregated systems) the evolution of the system between mergers (or until the system is relaxed) is a time-consuming phase with little scientific value. 
Since the time until the system segregates and violent relaxation is quenched is much shorter for a cold system than for a virial system, using cold initial conditions could, in the past, be a shortcut to the interesting part of the simulation. 
While these methods may be justified in some cases we are left to wonder if it is in general a valid approximation to the desired physical system.
Or for the case of mergers, what effect a free falling interaction, i.e. when using cold initial conditions, may have on impact parameters that a more gentle spiralling interaction, as in the case of warmer initial conditions, may not have.

\subsection{Violent Relaxation}
\citet{Lynden-Bell1967} attempted to explain the ``observed light distributions of elliptical galaxies'' and in doing so produced the first theory to describe the steady state resulting from a collisionless gravitational collapse.
In that pioneering work we find the first use of the term violent relaxation to describe the ``violently changing gravitational field of a newly formed galaxy''.
The fundamental premise of the theory is that the stars in a galactic model may be treated as a large set of independent, non-interacting harmonic oscillators.  
These oscillators are treated statistically and are expected to find a state of maximum entropy.
The weakness of the theory lies in the last statement.
During the collapse the system does not have enough time to explore the phase space and so will not generally come to equilibrium in the predicted state.

Since the work of \citet{Lynden-Bell1967} several other attempts have been made to extend, modify, and completely rework the theory of violent relaxation \citep[e.g.][]{Spergel1992,Nakamura2000,Treumann2013}.
In spite of these efforts, difficulties remain in constructing a theory which adequately describes the behavior of what may seem at first glance, a simple system \citep{Arad.Lynden-Bell2005,Arad2005}.

\subsection{Notation}
We recall that the virial temperature is $Q \equiv \lvert T/V\rvert$, where $T$ and $V$ are the kinetic and potential energies, respectively, and that a system in virial equilibrium has a $Q$ value of 0.5.  
Note that just because the energetics of the system is in equilibrium does not imply that the system as a whole is in equilibrium.
For example, a system with a $Q$ value of 0.5 can still be out of equilibrium if the system has a uniform density distribution (homogeneous sphere), as used in this paper, this is because the homogeneous sphere is not a solution to the Fokker--Planck equation.

We define the term fraction of virial (\fovmath) to be the current system's $Q$ value over the $Q$ value of a virialized system, or simply $2Q$.  
This definition conveniently results in a virialized system having a \fovmath = 1.  
We also define the term velocity multiplier, $\mathit{k}$, as the value the velocity is initially multiplied by to change the system from virial, that is: $k=v/v_{vir}$, $v_{vir}$ is the virial velocity of a particle.
So we find that initially 
\begin{align*}
\fovmath =& \, 2Q = 2\frac{\sum_{i}\frac{1}{2}m_i(\mathit{k}v_i)^2}{\sum_{i} V_i}\\ \bigskip
=& \, k^2\times 2 Q_{vir} = k^2.
\end{align*}

\section{Simulations}
\subsection{\textsc{AMUSE} }
Our simulations were run in the \textsc{AMUSE} software environment \citep{AMUSE2012}.  
\textsc{AMUSE} is a modular simulation platform which provides a set of simulation codes linked together through a \textsc{python} interface.
Different codes can be used on the same initial conditions, allowing for a fast, simple, and clear test of consistency between codes; \textsc{AMUSE}'s modular nature makes this easy to do, usually requiring a change to only two lines of code.
For example, we tested our simulations with three different N-body integrators, namely: \textsc{Hermite} \citep{Hut1995}, \textsc{PhiGRAPE} \citep{Harfst2007}, and \textsc{ph4} (McMillan in preparation).  
Again the \textsc{AMUSE} framework ensured the changes to the code were trivial, and by testing with different integrators we obtain an assurance that our results are not code-dependent, since all three codes gave similar results.  
When using the same set of initial conditions for example, plots of the half-mass radius versus time are nearly indistinguishable, and the number of bound particles at the end of the simulation never differ by more than 55 particles and on average differ by fewer than nine particles (less than 0.37 and 0.06 per cent of the total number of particles respectively).
We are now comfortable to assert that the results we present within this work are not the effect of a bug or a strange implementation found in one code, but represent the outcome of physical processes acting on our initial conditions.

All the data presented in this work were produced using \textsc{ph4}.
A parallel fourth-order \textsc{Hermite} integrator \textsc{ph4} can, and for us does, use GPUs to accelerate the computational work (this is accomplished through the use of the Sapporo library \citep[][B{\'e}dorf in preparation]{Gaburov2009, Gaburov2012}).  
We find it important to use a direct integrator for these simulations, as opposed to a tree code, because strong interactions play a role in the systems we aim to investigate.
In the analysis, we made extensive use of the group finding code hop \citep{Eisenstein1998}.
The runs were performed on the Little Green Machine, a local GPU cluster using NVIDIA GPUs.

\subsection{Initial Conditions}\label{sec:initial_conditions}
As this paper is focused on the effect initial conditions have on the resultant physical system we thought it only appropriate to explain exactly how the initial conditions presented within these pages were created.
We chose the initial conditions in the following way: a number of particles are distributed in a homogeneous sphere.
A homogeneous sphere is used in order to isolate the effects of violent relaxation which can become muddled when using more complex distributions.
The mass of the whole system is set to 1.0 N-body mass \citep{Heggie1986} and either the mass is divided equally amongst all star particles or, in order to study the effects of a more realistic mass function, a Salpeter mass function, having a slope of 2.35 \citep{Salpeter1955}, with a mass range N-body mass equivalence between 0.3 and 100 $M_{\odot}$ is applied or the mass is divided equally amongst all star particles.  
Each particle is given a velocity drawn randomly from a Gaussian distribution centered at zero, producing an isotropic velocity distribution.
If a black hole has been included, it is given a velocity of zero and placed at the center of the cluster.
Then the whole system is scaled to be in virial equilibrium. 
Finally, all unbound particles (particles with an energy~\textgreater ~0) are removed; this is the only time that particles are removed from the system.
These particles, along with their position and velocity, are saved to a file.
We repeat this procedure with different random initializations always requiring that the final number of objects bound to each system be the number of objects desired $\pm 5$ (never differing by more than 5).
Each set of initial conditions is produced four times, each with a different random realization of the particle positions to quantify the effects from initial position on the evolution of the system and to measure the statistical noise.

Before the start of the simulation, the velocities are scaled to whatever \fovmath is being investigated in that run, that is we multiply the velocity by $k$, the velocity multiplier.  
Using the same set of initial conditions for an entire set of runs ensures that the differences in each simulation are only due to the difference in velocity.
We use 21 values of $k$ (from 0.0 to 2.0 in 0.1 increments) to explore the effect of the \fov on the system.
Note that for the supervirial runs particles may be initially unbound, and in many of the subvirial cases particles become unbound after some time, but these particles are never removed from the simulation.

\section{Results and Discussion}
\label{sec:results}
\begin{table}
\caption{Outline of Simulations}
\label{tab:initial_conditions}
\begin{tabular}{ccccc}
No. of runs & No. bound & Density & $M_{\rm bh}$ & Mass function\\
&particles&&[N-body mass]&\\ \hline
$4 \times 21$&15210&Uniform&0&Equal mass\\
$4 \times 21$&15210&Uniform&0&Salpeter\\
$4 \times 21$&15210&Uniform&0.02&Equal mass\\
$4 \times 21$&15210&Uniform&0.02&Salpeter\\
$21$&15210&Uniform&0.05&Salpeter\\
$21$&15210&Uniform&0.10&Salpeter\\
$21$&15210&Plummer&0.02&Salpeter\\
$21$&15210&King ($\omega = 6$)&0.02&Salpeter\\
$21$&2048&Uniform&0.02&Salpeter\\
$21$&4096&Uniform&0.02&Salpeter\\
$21$&8192&Uniform&0.02&Salpeter\\
$7$&131072&Uniform&0.05&Salpeter\\
\end{tabular}
\end{table}
The simulations we ran are described in Table~\ref{tab:initial_conditions}.  
Column 1 gives the number of runs performed with each set of initial conditions.
Each set of initial conditions (save the last set) are run with 21 different \fovmath, ranging from 0.0 to 4.0 ($Q=0.0-2.0$); the first four sets are simulated with four different random realizations of the particle distribution in order to reduce statistical error.
The \fov is chosen such that the velocity multiplier, $k$, is equally spaced in 0.1 intervals, i.e. 0.0, 0.1, 0.2,\ldots, 1.9, 2.0.   
Column 2 of Table~\ref{tab:initial_conditions} gives the number of bound particles at the start of each simulation (see Section ~\ref{sec:initial_conditions} for more information about our initial conditions).
The Salpeter mass function was generated with an N-body mass unit equivalent to 0.3-100 $M_{\odot}$.
All simulations are run for a minimum of 10 N-body time units \citep{Heggie1986} with a data output rate of 50 snapshots per N-body time.
We use a softening length, $\epsilon$, such that $\epsilon ^2 = 10^{-8}$ for all simulations except for the simulations with 131,072 bound particles where we use an $\epsilon ^2 = 10^{-16}$ to be sure we capture the detail of the interactions.
In total we ran 490 simulations. 
\subsection{Escape Fraction}
\begin{figure}
  \centering
  \subfloat[Equal mass particles]{%
    \label{fig:bound_FOV_sub1}
    \includegraphics[width=0.5\textwidth]{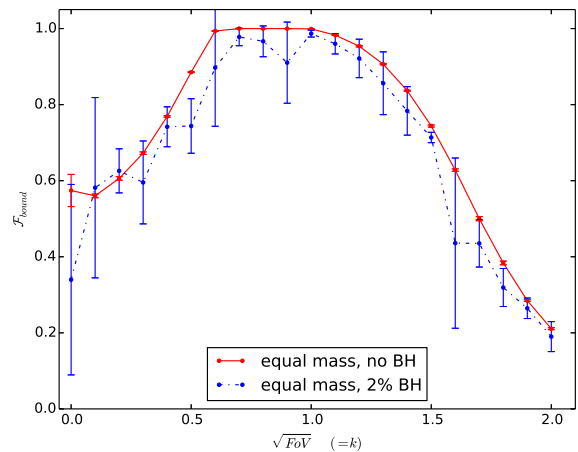}
  }\\
  \subfloat[Salpeter mass function]{%
    \label{fig:bound_FOV_sub2}
    \includegraphics[width=0.5\textwidth]{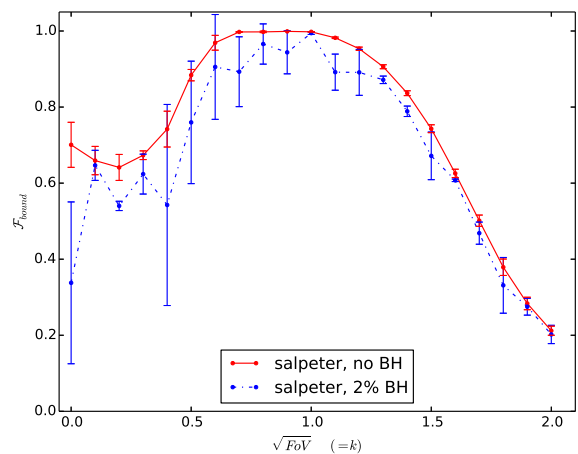}
  }
  \caption{\small{The fraction of objects remaining bound to the system versus the \fovmath.  
           The solid, red line is for simulations run without a black hole, and the dashed, blue line is for simulations with a black hole of 2 per cent of the total mass of the system.  
	   The error bars represent one standard deviation.} }
  \label{fig:bound_FOV}
\end{figure}
Figure~\ref{fig:bound_FOV} is a plot of \fov versus the fraction of objects that remain bound to the system after 10 N-body times.  
Each data point is an average of at least four runs, and the bars indicate one standard deviation, i.e. a measure of the spread, not the error.  
Figure~\ref{fig:bound_FOV_sub1} shows the results of simulations, with equal mass particles (save the black hole) both with and without a black hole.  
The black hole, when present, contains 2 per cent of the total mass of the system.  
Figure~\ref{fig:bound_FOV_sub2} is a plot of the same simulations with the exception that the objects' masses are chosen from a Salpeter mass function; again the cases with and without a black hole are shown and the error bars represent one standard deviation.
Though not shown we also ran simulations using a black hole with 5 and 10 per cent of the cluster mass.
These simulations showed a similar shape to the curves shown in Figure~\ref{fig:bound_FOV} but generally with fewer particles remaining bound as the mass of the black hole was increased. 

We note an uptick in retained number of particles with a \fov of 0.0 versus 0.01 for the equal mass systems without a black hole, and a \fov of 0.0 versus 0.04 for systems with a Salpeter mass function without a black hole.  
To verify that the uptick was not simply an artifact of our four standard realizations, 21 more runs with different random realizations were performed (for a total of 25 realizations) with a $\fov = 0.0$, no black hole, and Salpeter mass function.
The results of all 25 realizations are plotted for that point in Figure~\ref{fig:bound_FOV}.

A possible interpretation for such an uptick is that when the system begins cold ($\fov = 0.0$) there is no radial motion so the particles follow a nearly free-fall trajectory towards the center of mass and so spend the least amount of time in the very high density of the collapse.  
(The reduced time spent in the highest density of the collapse for cold systems can be seen by comparing the Lagrangian radii in both panels of Figure~\ref{fig:FOV_lagrangian}.)
However, as the \fov increases there is increasing radial velocity leading to an in fall trajectory which is more spiral-like than free-fall-like.
With a low but non-zero \fov still leading to a very dense collapse and the particles spending more time near the center of mass at the time of deepest collapse the probability of interactions increases resulting in a higher likelihood for scattering events.  
When a black hole is added a free-fall path aimed directly at the center is almost a guarantee for a strong scatting event, as can be seen in the cases with a black hole (dashed, blue line) and a $\fov = 0.0$.

In the cases with a mass function the fraction of \emph{mass} retained by the system is always greater than or equal to the fraction of particles retained indicating that we keep the more massive particles preferentially and thus tend to lose low-mass particles as expected.  

We find in both panels of Figure~\ref{fig:bound_FOV} the effect of including a black hole is to, in general, reduce the number of bodies remaining bound to the system, as well as to produce more noise in the measurement.
We can understand this by noting that interactions leading to ejections between particles with similar masses only, as is the case when a system does not possess a black hole, produce the loss of bodies as seen in the red lines in Figure~\ref{fig:bound_FOV}.
Introducing a black hole to a system does not change the number of interactions between particles with similar masses and so ejection rates between such particles remain similar to the case without a black hole.
However, as the black hole interacts with particles there is the additional case of large mass ratio interactions leading to ejections from the system over the similar mass ejection rate baseline.
Thus, the reason the systems without a black hole tend to provide an upper limit on the number of particles remaining bound to the system is due to the addition of a strong scatter in the systems with a black hole while not changing, very much, the probability of smaller mass ratio scattering events.
The additional noise found in these measurements of systems with a black hole is the result of the scattering by the black hole being sensitive to the exact nature of the interaction and thus to the random realization of the particles initial positions and their relative velocity.

\citet{Proszkow2009ApJS} and \citet{Adams2006} measured the number of objects that remain bound after 10~Myr for different \fov but include additional effects such as primordial mass segregation, a static gas potential, and gas removal.
The difference in the shape of the fraction remaining bound in \citet{Proszkow2009ApJS} is likely due to their static gas potential and analytic gas removal, resulting in a change of the potential energy of the system.
It seems this would be similar to a change in the initial \fovmath, though it is not clear that such a simple substitution would be correct.
For instance, if the gas is removed from the system before or even shortly after the collapse (see Figure~\ref{fig:FOV_lagrangian} and Section ~\ref{sec:mass_seg} for a description of what is meant by collapse) the system's evolution will be different than if the gas is removed after the system has relaxed and has reached, or very nearly reached, virial equilibrium.
However, without performing simulations that examine this parameter space, preferably modeling a coupled evolution of the stellar and gas dynamics, we cannot be sure how, or if, the initial virial temperature can be used as a proxy for gas removal.
Moreover, the nature of gas removal from clusters (e.g. the amount removed, the age of the cluster when it is removed, the dependence of gas removal on cluster mass, et cetera) is still being investigated \citep[see e.g.][]{Dale2014,Pelupessy2012}.

\subsection{Mass Segregation}
\label {sec:mass_seg}
\subsubsection{Bound versus Unbound:a Cautionary Note}
\begin{figure}
  \centering
  \subfloat[Includes all particles (bound and unbound)]{%
    \label{fig:all_lagrangian_sub1}
    \includegraphics[width=0.5\textwidth]{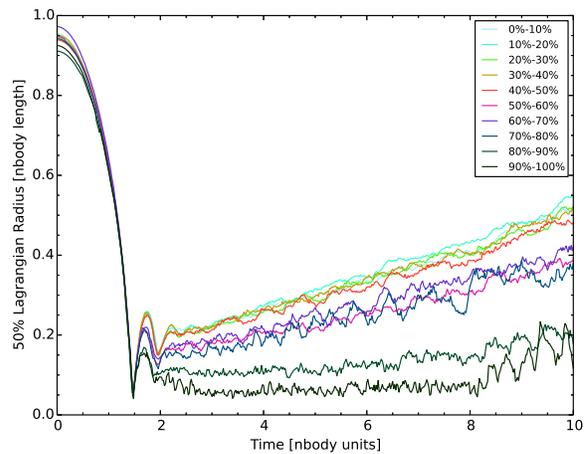}
  }\\
  \subfloat[Includes only the bound particles]{%
    \label{fig:bound_lagrangian_sub2}
    \includegraphics[width=0.5\textwidth]{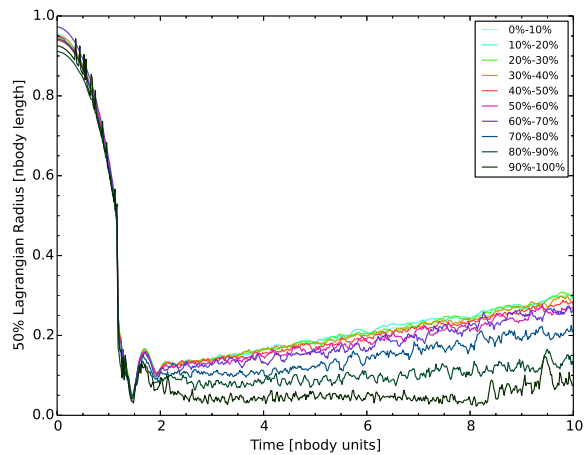}
  }
  \caption{\small{
	   The 50 per cent Lagrangian radius (or half-mass radius) for 10 per cent mass bins of a system with a \fovmath=0.0.
           Each bin contains 10 per cent of the mass and in general the upper lines represent lower mass objects while the lower lines represent higher mass objects.
           The top panel has all the particles which were originally in the system plotted regardless of whether they remain bound to the system.
           The bottom panel has only the particles which are bound to the system at that given time.
           Each different decade of mass is clearly identifiable and more spread out in the top plot, whereas the data are more compressed and mixed in the bottom plot.} } 
  \label{fig:all_vs_bound_lagrangian}
\end{figure}

In Figure~\ref{fig:all_vs_bound_lagrangian}, we plot the 50 per cent Lagrangian radii, i.e. the half-mass radius, using data from a simulation with an initial \fovmath=0.0 and a black hole containing 2 per cent of the total mass.  
First, the system collapses in approximately a free-fall time to a depth which is often given as $\approx N^{-1/3}$ for N-body simulations.
In Sections ~\ref{sec:timescales} and ~\ref{sec:depth} we discuss the time and depth of the first collapse in more detail.
Next, the system rebounds and undergoes a second collapse which is not as deep as the first, similar to a damped oscillator.

Before plotting Figure~\ref{fig:all_vs_bound_lagrangian} we divide the objects, excluding the black hole, into bins of 10 per cent of the total mass, thus the more massive bins have fewer particles. 
After the collapse the bins with the most massive objects tend towards smaller radii (bottom of the plot), and conversely bins with the least massive objects can be found with larger half-mass radii.
For example, in both panels the bottom line contains the most massive objects which collectively comprise a total of 10 per cent of the system mass, and while the mass represented in each bin is the same it will represent different numbers of objects. 

The top panel, Figure~\ref{fig:all_lagrangian_sub1}, shows the half-mass radii of the system when including both bound and unbound particles; whereas the bottom panel, ~\ref{fig:bound_lagrangian_sub2}, shows the half-mass radius of the system including only the particles bound to the system at each snapshot.
The distinction is important particularly for the simulations with low values of the \fov which lose a large fraction of the initial objects.
The top panel of Figure~\ref{fig:all_vs_bound_lagrangian} is in good agreement with the results from \citet{McMillan2012}.

When plotting all particles, as compared to only the bound particles, the system appears to have a larger half-mass radius due to the unbound particles tending to be further away from the system and thus increasing the apparent half-mass radius.  
This is particularly noticeable in the lower mass bins since they are preferentially lost.

However, by taking both bound and unbound particles into account for the analysis the expansion of the cluster appears to be much faster than when only the bound particles are plotted.  
This would likely lead to a wrong estimate of the evaporation time-scale for the system (presumably other measures of system-wide parameters would be similarly affected).
Furthermore, the cluster appears mass segregated even in the lower mass bins, but in fact the selective expulsion of low-mass stars is mimicking mass segregation for these stars. 
The bottom panel makes clear that the (bound) cluster expands much more slowly and the mass segregation is only significant for the highest mass bins.

There appears to be more mass segregation when all particles are plotted.   
For example, in Figure~\ref{fig:all_lagrangian_sub1} the 40 per cent of the mass contained in the least massive particles (i.e. the top four lines in the plot) is not segregated but segregation is noticeable between the most massive 40 per cent and the 50--60 per cent range, and each decade of mass after that.  
Whereas for Figure~\ref{fig:bound_lagrangian_sub2} there is no appreciable segregation in the 0 to 70 per cent range of the mass.
The degree of segregation between the various decades of mass is more pronounced when plotting all particles, i.e. the differences between the half-mass radius for the top 10 per cent of the mass (the very dark green line in the plots) and the decade below that (the green line) are larger when plotting all particles (the top panel).
These differences would lead to a much different conclusion about the nature of an observed or modeled cluster.
Since most objects which become unbound from a system are likely to be long gone at the time of observation, the plots with only bound stars demonstrate a more correct system.

Moreover, without making this distinction the apparent results from the simulation do not reflect the dynamics occurring in the system, since unbound particles which, in time, have almost no impact on the dynamics are still being analyzed as if they were dynamically important.
Unless noted otherwise, we shall only use the bound objects at each snapshot for further analysis.
\subsubsection{Effect of the Initial \fov on Mass Segregation}
\begin{figure}
  \centering
  \subfloat[\fov = 1.0]{
    \label{fig:FOV_1_lagrangian_sub1}
    \includegraphics[width=0.5\textwidth]{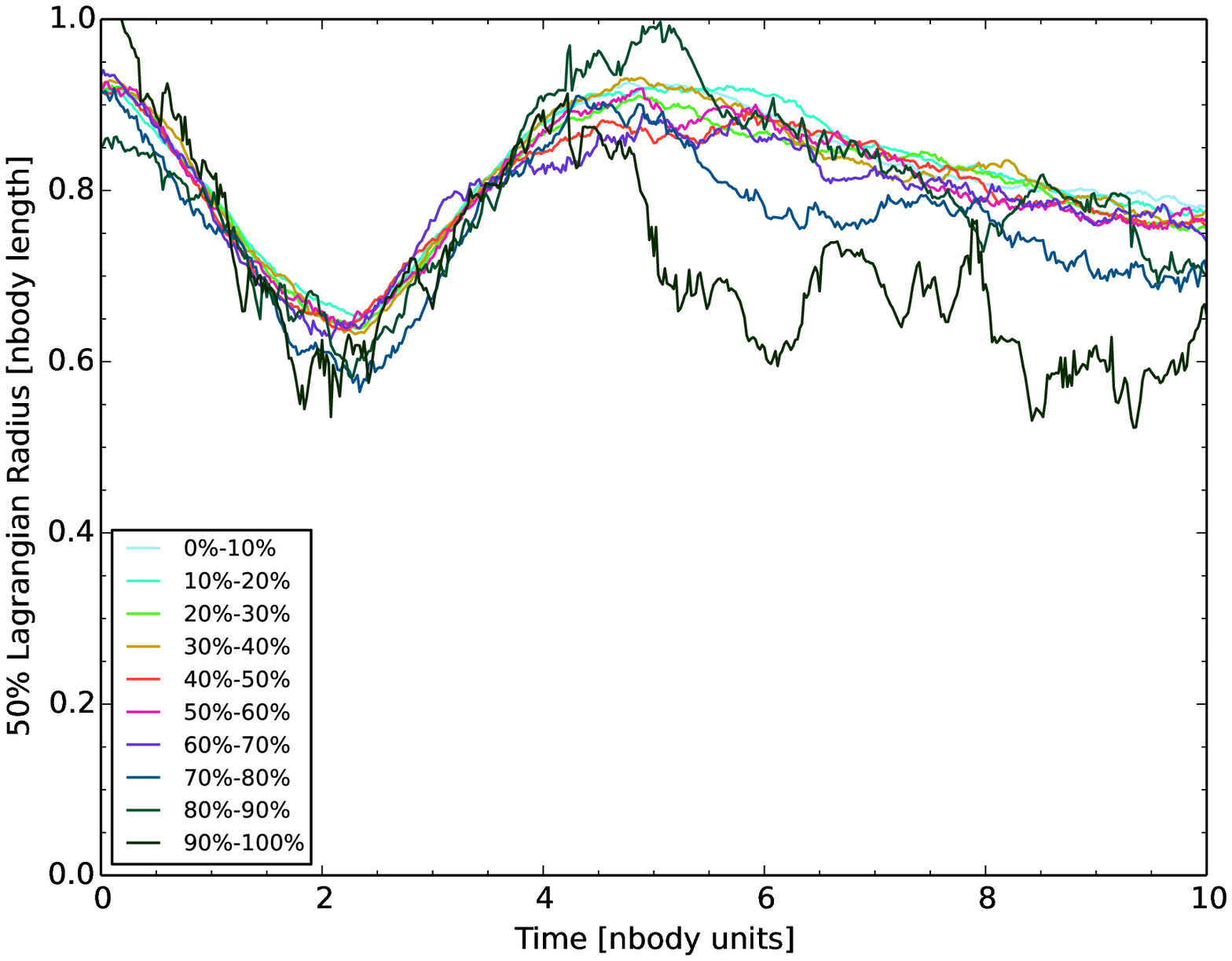}
  }\\
  \subfloat[\fov = 0.0]{
    \label{fig:FOV_0_lagrangian_sub2}
    \includegraphics[width=0.5\textwidth]{lagrangian_radius_bound}
  }
  \caption{\small{The 50 per cent Lagrangian radius plotted for 10 per cent mass bins.
           The upper plot shows the half-mass radii for a system with \fovmath=1.0 (virial) and the lower plot for a system with \fovmath=0.0 (completely cold).
           } }
  \label{fig:FOV_lagrangian}
\end{figure}

In Figure~\ref{fig:FOV_lagrangian}, we plot the half-mass radii, just as we did in Figure~\ref{fig:bound_lagrangian_sub2}.  
The upper panel, Figure~\ref{fig:FOV_1_lagrangian_sub1}, shows the half-mass radii for a system with an initial \fov of 1.0 (virial), while the system in the lower panel, ~\ref{fig:FOV_0_lagrangian_sub2}, had an initial \fov of 0.0 (cold).
Just as before, the very dark green line represents the most massive particles which comprise 10 per cent of the mass, and the green line above that represents the second set of most massive particles which comprise the next 10 per cent of the mass.

The depth of collapse (i.e. the minimum radius of the system during collapse) is often given as $R_{min}\approx N^{-1/3}$.
We find this relationship to only hold for the case where the initial \fovmath=0.0, see Figure~\ref{fig:depth} and Section ~\ref{sec:depth} for the better fit we find for different initial \fovmath.
The depth of collapse becomes deeper when the initial \fov is lower (also see Figure~\ref{fig:depth}).

Segregation begins during the collapse for both systems and is realized at the deepest collapse.
This fast mass segregation has been examined by \citet{Allison2009} and \citet{Allison2010}, and observed in other simulations (e.g. \citet{Geller2013}).
In the case of the cold system the bounce occurs at $\approx 1.8$ N-body times, whereas for the virial case it requires $\approx 5$ N-body times.  
The virial case takes longer to segregate due to its longer time until collapse, as seen in the insert of Figure~\ref{fig:timeScales}.
The increase in density found at the depth of collapse is what allows the segregation to occur so quickly, and since the deeper the collapse the higher the density so the faster segregation can occur.

We observe, as mentioned above, the collapse is much deeper and shorter in the cold case, Figure~\ref{fig:FOV_0_lagrangian_sub2}, than in the virial case, Figure~\ref{fig:FOV_1_lagrangian_sub1}, but we find it to have a very different segregation signature.
That is, the difference in how the mass is segregated, not so much in the degree of segregation but rather in the degree of segregation between the different mass ranges.
This is an example where the attempt to shortcut the cost of evolution using a cold system is clearly seen.

Many of the system properties change as a function of the \fovmath.
Cold initial conditions are sometimes used to more quickly reach a relaxed system (see Figure~\ref{fig:timeScales} for evidence of faster relaxation for cold systems).
In doing so, the implicit assumption is that a relaxed cluster has no memory of the initial \fov but clearly this is not the case; clusters with different initial \fov result in clusters with different relaxed radii and number of bound particles, for example.
Some, but not all, of these differences might be resolved by scaling of the initial and final systems, though this would likely come at the expense of faster relaxation.

Moreover, we cannot suggest a way to scale the segregation signature and without scaling it the system will always remain physically distinct.
This might however provide an interesting way to diagnose the initial \fov of observed young clusters, though more work would be required in understanding the impact the initial \fov has on the segregation signature (see Section \ref{R136}).

It should be noted that the collapse seen in the system with $\fovmath=1$ (Figure~\ref{fig:FOV_1_lagrangian_sub1}) is not due to non-equilibrium in the global energetics of the system, but rather due to the spatial and velocity distributions of the particles not being in a relaxed state (i.e. not a solution to the Fokker--Planck equation).

\subsection{Time-scales}\label{sec:timescales}
\begin{figure}
  \includegraphics[width=0.5\textwidth]{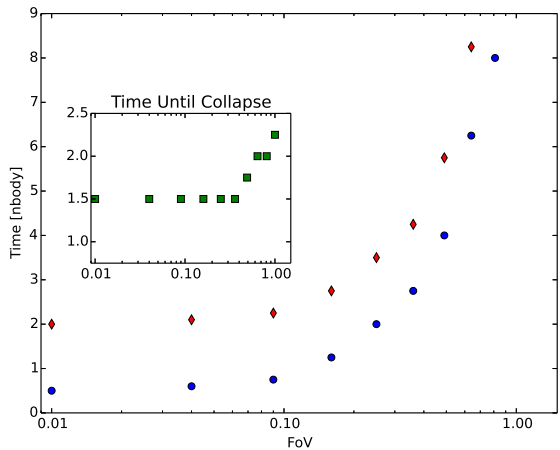}
  \caption{\small{The green squares, of the inset plot, mark the time from the beginning of the simulation until collapse.  The blue circles indicate the time from the collapse until the end of the bounce.  The red diamonds mark the length of time from the beginning of the simulation until the end of the bounce for each \fovmath; in other words, the minimum time required to simulate in order to reach a mass segregated and relaxed system.} }
  \label{fig:timeScales}
\end{figure}

The inset of Figure~\ref{fig:timeScales} is a plot of the time until the deepest collapse of the system, and the blue circles plot the time between the moment of deepest collapse of the half-mass radius until the end of the bounce for each \fovmath. 
The red diamonds mark the time from the beginning of the simulation until the end of the bounce.

We assume that  after the bounce the effect of violent relaxation is minimal and the system enters a new regime where two-body relaxation begins to dominate.
The time required to reach a virially relaxed state increases as the \fov increases, this should be expected since this time is simply the sum of the time until collapse (inset in Figure~\ref{fig:timeScales}) and the time from collapse until rebound (the blue circles in Figure~\ref{fig:timeScales}) both of which increase with \fovmath.

The red diamonds in Figure~\ref{fig:timeScales}, provide evidence that warm initial conditions in fact do require simulating for more crossing times than cold ones.
Note that it takes more than five times longer for the initially virialized case than for the initially cold case to reach the end of the bounce (\textgreater 10 N-body times compared to 2 N-body times).
The end of the bounce for the virial case ($\fovmath=1$) is not seen before the 10 N-body times for which we ran these simulations.

The inset in Figure~\ref{fig:timeScales} shows the time until the system reaches the deepest point of collapse, or $R_{min}$.  
We know that the free-fall time-scale, which is the time for collapse of a homologous contraction, is
\begin{equation}
 \tau_{\mathtt{FF}} = \sqrt{\frac{3\pi}{32G\rho}} \; .
\end{equation}
Keeping with our use of N-body units, $G=M=1$, thus $\rho = \frac{3}{4\pi R^3}$ and our equation reduces to
\begin{equation}
 \tau_{\mathtt{FF}} = \frac{\pi}{2}\sqrt{\frac{{R^3}}{2}}.
\end{equation}
At the beginning of the simulations, we measure the most distant particle to be $\approx1.2$ N-body lengths from the center of mass of the system, using that value for the radius we find a constant value for the time of collapse in our simulations to be
\begin{equation}
 \tau_{\mathtt{collapse}} \approx 1.46.
\end{equation}
This value is close to what is plotted in the inset of Figure~\ref{fig:timeScales} for a \fov between 0.0 and 0.36.  
However, we show it is not valid to assume a free-fall time-scale as the relevant time-scale for collapse in a system with an initial \fov \textgreater 0.36 ($Q > 0.18$). 

\subsection{Minimum Cluster Radius}\label{sec:radius}
\label{sec:depth}
\begin{figure}
  \includegraphics[width=0.5\textwidth]{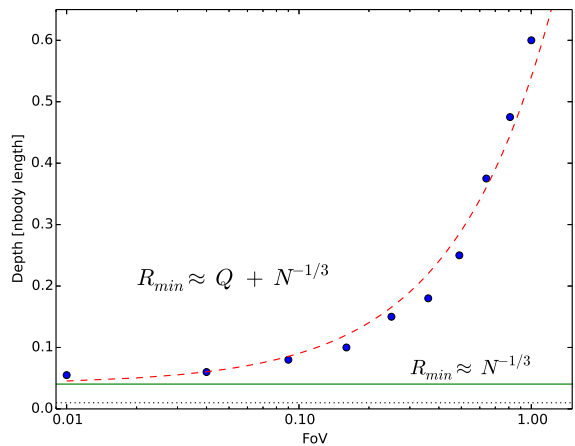}
  \caption{\small{The half-mass radius at the point of deepest collapse versus the \fovmath.  The red dashed line is the fit we propose with a $Q$ dependency, the green solid line is the theoretically predicted value, and the black dotted line is the softening length.} }
  \label{fig:depth}
\end{figure}

In Figure~\ref{fig:depth}, we show the half-mass radius at the point of deepest collapse, i.e. the minimum radius during the collapse, versus the \fovmath. 
In this figure, we demonstrate the dependence of the depth of the collapse as a function of \fovmath.

The depth of the collapse, $R_{min}$ of collapse, is often given as $\approx N^{-1/3}$ \citep{Aarseth1988}, where $N$ is the number of particles.
We recover a value very close to this for the case of a cold collapse finding a difference of only 0.01 N-body lengths.
However, as we show in Figure~\ref{fig:depth}, and can also be seen in Figure~\ref{fig:FOV_lagrangian}, the depth of collapse is also dependent on the \fovmath.
In our experiments, we hold $N$ constant and change \fov and we find that as the system becomes more virial the collapse becomes less deep, that is $R_{min}$ becomes larger.
We find that 
\begin{equation}
 R_{min} \approx \frac{1}{2} \times \text{\fovmath} + N^{(-1/3)}
\end{equation}
provides a good fit to our data, and is a substantially better approximation for $R_{min}$ in non-cold systems.

Recall from our definition of \fov that $\frac{\fovmath}{2}$ is equal to $Q$.
So finally we propose that the minimum radius of collapse is dependent not only on $N$ but also on the virial temperature in the following way:
\begin{equation}
 R_{min} \approx Q + N^{(-1/3)},
\end{equation}
where $Q \equiv \lvert T/V\rvert$ and $N$ is the number of particles being simulated.

\subsection{Observables}
\begin{figure}
  \centering
  \subfloat[Core Radius]{%
    \label{fig:no_bh_core_radius_map}
    \includegraphics[width=0.47\textwidth]{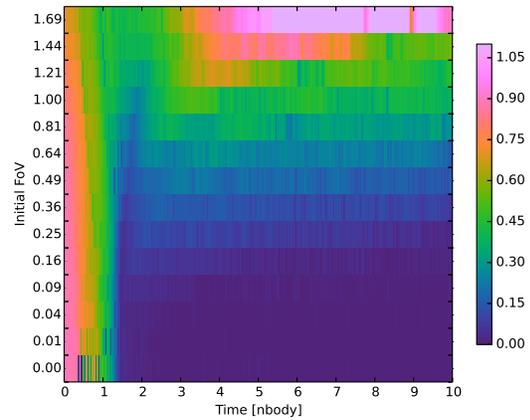}
  }\\
  \subfloat[Slope of the Density Distribution]{%
    \label{fig:no_bh_density_map}
    \includegraphics[width=0.47\textwidth]{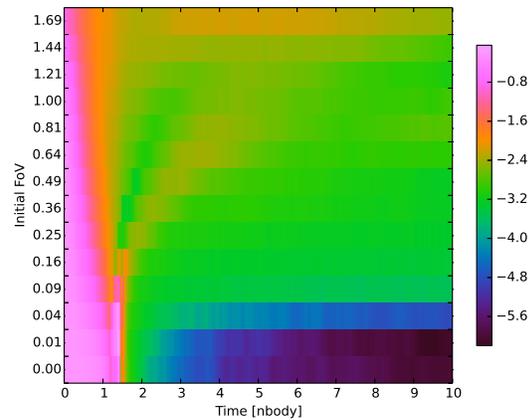}
  }\\
  \subfloat[Mass Segregation]{%
    \label{fig:no_bh_mass_seg_map}
    \includegraphics[width=0.47\textwidth]{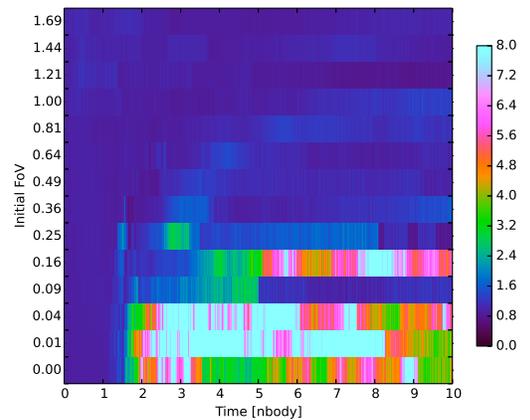}
  }
  \caption{\small{Observable quantities plotted against initial \fov and time for a system without a black hole.
           Top: colors denote the core radius in N-body units.  Middle: colors denote density distribution slope.  Bottom: colors denote mass segregation ratio.
           } }
  \label{fig:no_black_hole_observables}
\end{figure}

\begin{figure}
  \centering
  \subfloat[Core Radius]{%
    \label{fig:core_radius_map}
    \includegraphics[width=0.47\textwidth]{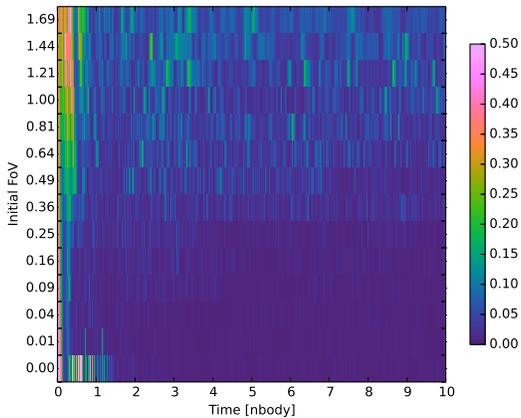}
  }\\
  \subfloat[Slope of the Density Distribution]{%
    \label{fig:density_map}
    \includegraphics[width=0.47\textwidth]{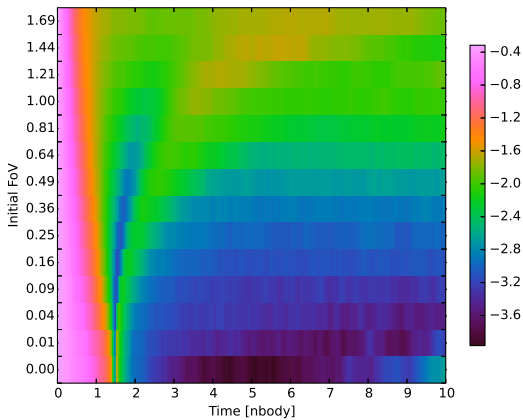}
  }\\
  \subfloat[Mass Segregation]{%
    \label{fig:mass_seg_map}
    \includegraphics[width=0.47\textwidth]{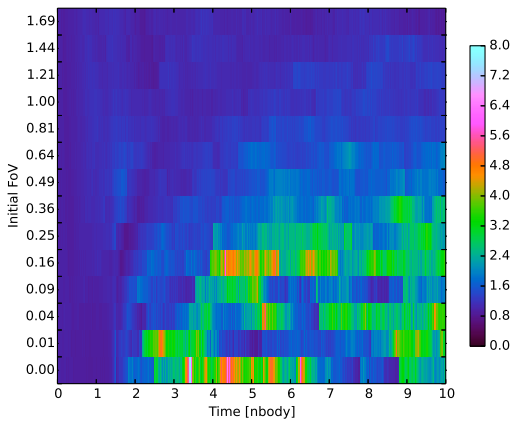}
  }
  \caption{\small{Observable quantities plotted against initial \fov and time for a system with black hole containing 2 per cent of the cluster mass.
           Top: colors denote the core radius in N-body units.  Middle: colors denote density distribution slope.  Bottom: colors denote mass segregation ratio.
           } }
  \label{fig:black_hole_observables}
\end{figure}

In Figures ~\ref{fig:no_black_hole_observables} and ~\ref{fig:black_hole_observables}, we provide plots of three observable parameters: the core radius, the slope of the density distribution, and the mass segregation ratio.
We calculate the core radius by following \citet{Casertano1985} with a density weighting factor of 2.
To measure the slope of the density distribution, we perform a linear least-squares fit of the density and radial distance from the center of the cluster in log--log space.
In measuring both the core radius and density distribution we determine the local density using hop \citep{Eisenstein1998} with a 7 neighbor particle radius.
The mass segregation ratio is calculated using the minimum spanning tree method described in \citet{Allison2009MST}.
We use the 20 most massive particles to construct the ``massive'' tree and 50 different sets of random particles to construct the ``random'' trees.
Figure~\ref{fig:no_black_hole_observables} shows data from simulations with a Salpeter mass function, with particles initially distributed in a homogeneous sphere, and no black hole, while Figure~\ref{fig:black_hole_observables} has a similar set of initial conditions with the addition of a black hole containing 2 per cent of the cluster mass.

We note several regimes in the plots: the first is at early times regardless of the \fov (the far left of the plots) there is a relatively large core radius, flat density distribution, and a small degree of mass segregation.
This of course is due to the initial conditions.

The second regime we note is the lower-right quadrant (small \fov and late times) where the systems have had time to relax.
Here we find the smallest core radii, the most extreme density distribution, and the highest degree of mass segregation.  
By mapping these quantities from an observed young cluster to Figures ~\ref{fig:no_black_hole_observables} and ~\ref{fig:black_hole_observables} along with other derived properties of a cluster (e.g. minimum age), constraints can be placed on the initial conditions of the system.
Additionally, the plots can be used to determine what range of \fov would be ideal to use in the initial conditions for a simulation which aims to reproduce a physical system or investigating a phenomenon in clusters with a particular observable parameter.

\section{Application to an observed cluster: R136}
\label{R136}
R136 is in the center of NGC 2070 (30 Doradus), which is in the Tarantula nebula, a young star cluster in the Large Magellanic Cloud (LMC).  
This region is the subject of many observations including two surveys: The VLT-FLAMES Tarantula Survey \citep{VLT-FLAMES2011} and the Hubble Tarantula Treasury Project \citep{HTTP2013}.
In the following, we attempt to simulate R136 as an isolated cluster in order to constrain the initial \fov and other properties.
For this purpose, we performed an additional set of simulations with initial conditions like the second row in Table~\ref{tab:initial_conditions}: 15,210 bound particles, no black hole, a Salpeter mass function, with particles distributed in a homogeneous sphere.
However, we run these simulations for 20 N-body times, producing 1000 snapshots for each simulation.

\subsection{Observed Parameters}
\citet{Hunter1995} found a core radius for R136 of 0.02~pc, a value that was refined to $0.025\pm 0.004$~pc by \citet{Andersen2009}.
The methods used to determine the core radius in \citet{Hunter1995} are disputed for example by \citet{Brandl1996}, who found core radii as a function of stellar mass cutoff ranging from $\approx 0.038$~to~0.3~pc for high- to low-mass cutoffs, respectively.
Other values for the core radius that have been proposed include 0.063~pc \citep{Campbell1992}, 0.1 and 0.15~pc \citep[using different filters,][]{deMarchi1993}, 0.2~pc \citep{Moffat1985}, 0.24~pc \citep{Malumuth1994}, and 0.33~pc by both \citet{Meylan1993} and \citet{Mackey2003}, though \citeauthor{Mackey2003} state that due to crowding in their images their value represents an upper limit.

\citet{Selman1999} provide us with a fit to the density profile with a single power law with an exponent of $-2.85$.  
There seems to be a much stronger consensus about the value of this observable in the literature and so we will use -2.85 with a spread similar to the range found in other works [section 3.3 in \citet{Selman1999} provides a good overview].

\citet{Sabbi2012} found that R136 likely started forming stars $\approx2$~Myr ago and was still active up to $\approx1$~Myr ago.
There are other, older age estimates for the cluster (e.g. \citet{Brandl1996} favor an age of $\approx 3.5$~Myr), but since \citet{Sabbi2012} differentiate between R136 and a separate clump to the northeast of R136, which is older and seems to be included in previous age estimates, we choose to use their value.
The young age of this cluster is ideal for comparing to our simulations since two-body relaxation has not yet had a strong effect on the system.

Finally, \citet{Henault2012} offer an in-depth analysis of the current virial state of R136.
After accounting for the rotation velocity and angle, variable stars, and binaries (see \citet{Gieles2010} for more about the impact of binaries on the virial state of young clusters) \citet{Henault2012} find that R136 is in virial equilibrium.

\subsection{From N-body to Physical Units}
So far we have shown our results in N-body units \citep{Heggie1986} however if we are to compare the results to R136 we will need to convert to physical units.
When the initial conditions (i.e. the physical scales) are known, this conversion is straightforward.
For example, by taking the ratio of the observed virial radius to the measured simulated virial radius, and the ratio of the observed mass of the cluster to the measured mass of the simulated cluster, and setting the gravitational constant to unity a complete converter from N-body to physical units is formed.
This converter can then be applied to each snapshot.

However, because we are attempting to constrain the initial conditions we cannot make an assumption about the initial physical scales (i.e. the mass and radius) of the system.
Moreover, as we are comparing our results to a known physical system for which we are not certain of the age in crossing times, i.e. N-body time units, we cannot assume that any particular snapshot is the one which represents the observed state.
Thus, we are forced to evaluate each snapshot as if it were the one which corresponds to the observed state and thus each snapshot must have its own conversion to physical units.

Our conversion from N-body units to physical units is accomplished in the following way: 
for every snapshot, we measure the half-mass radius of the bound particles, then, to simulate an observation which is seen in projection, we select all (bound and unbound) particles within a cylinder with a radius equal to the measured half-mass radius.  
Next, we measure the mass of all of the particles within that cylinder.
The final measurement we make is of the virial radius of the system.
This measurement must be done carefully since often these systems are out of virial equilibrium, so we use a definition based on the potential energy
\begin{equation*}
R_{vir}=-GM^{2}/(2V),
\end{equation*}
where $V$ is the potential energy.

Still all of these measurements are in N-body units, to convert we use a virial radius of 2.89~pc \citep{SPZ2010} and a total cluster mass of $10^5 M_{\odot}$ \citep{Andersen2009}.
We simply take the ratio of the observed virial radius to the simulated virial radius, and the total observed cluster mass to twice the simulated measure of the half-mass. 
These values along with setting $G=1$ make a complete unit conversion possible.
This procedure is repeated for every snapshot, in this case 1000 snapshots for each value of the initial \fovmath.

Since each snapshot has a different conversion factor there is counterintuitive behavior in some of the measurements.
As said, in many applications a simple (constant) conversion from N-body time units to physical age is possible, but since our snapshots are produced at fixed intervals of N-body time, and each one has a different conversion factor, the apparent age does not increase linearly, and sometimes may even decrease.
For example, if the radius of the cluster expands fast enough the time conversion factor may decrease more quickly than the time in N-body units has increased.

\begin{figure}
  \includegraphics[width=0.48\textwidth]{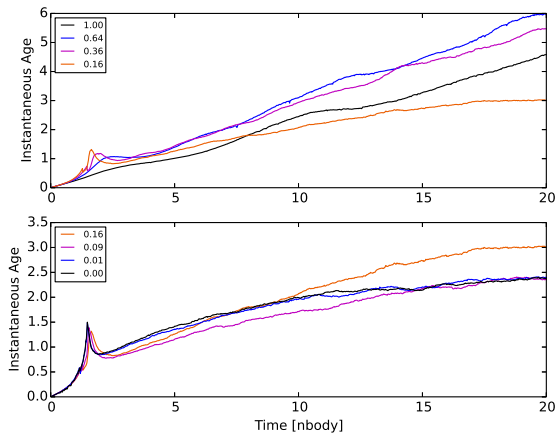}
  \caption{\small{The physical age at each snapshot using our converter from N-body units.
		 Each line corresponds to a different initial \fov with 0.16 plot in both panels.}}
  \label{fig:nbody_to_physical_time}
\end{figure}

In Figure~\ref{fig:nbody_to_physical_time}, we plot the calculated instantaneous age of each snapshot versus the N-body time using the conversion described above.
The bottom panel of Figure~\ref{fig:nbody_to_physical_time} are plots of the age for initial \fov of 0.0 to 0.16 while in the top panel the plots for 0.16 to 1.21.

The prominent spike in many of the simulations around 1.5-2 N-body times is due to the collapse of the system.
During the collapse the simulated half-mass radius is decreasing very rapidly while the simulated mass interior to the projected half-mass radius is remaining constant so the physical time evolved per snapshot becomes very large.
Another way to word it is that as the system collapses the number of crossing times per snapshot is increasing.

Again we would like to point out, as we did in Section \ref{sec:mass_seg}, the importance of discerning when it is appropriate to use only the bound particles or all (bound and unbound) particles.
To demonstrate this point, we performed the conversion as described above but using the bound and unbound particles to make the measure of the radius (instead of using the particles in the selection cylinder as we did for this analysis).
When making this measurement on all the simulated particles we obtained different results, but most strikingly we found that the instantaneous age of each snapshot began to monotonically \emph{decline} after a few N-body times.
This is due to the virial radius growing too large too quickly, because of the escaping unbound particles.
Clearly such behavior is unphysical, since it would imply that even with an infinitely long simulation the physical age would not increase beyond a certain point, but without such a  plot it might not be obvious that something was amiss.

\subsection{Initial Virial Temperature of R136}
\begin{figure*}
  \subfloat[Core Radius]{%
    \label{fig:cylinder_core_radius_clumpy_no_black_hole}
    \includegraphics[width=0.48\textwidth]{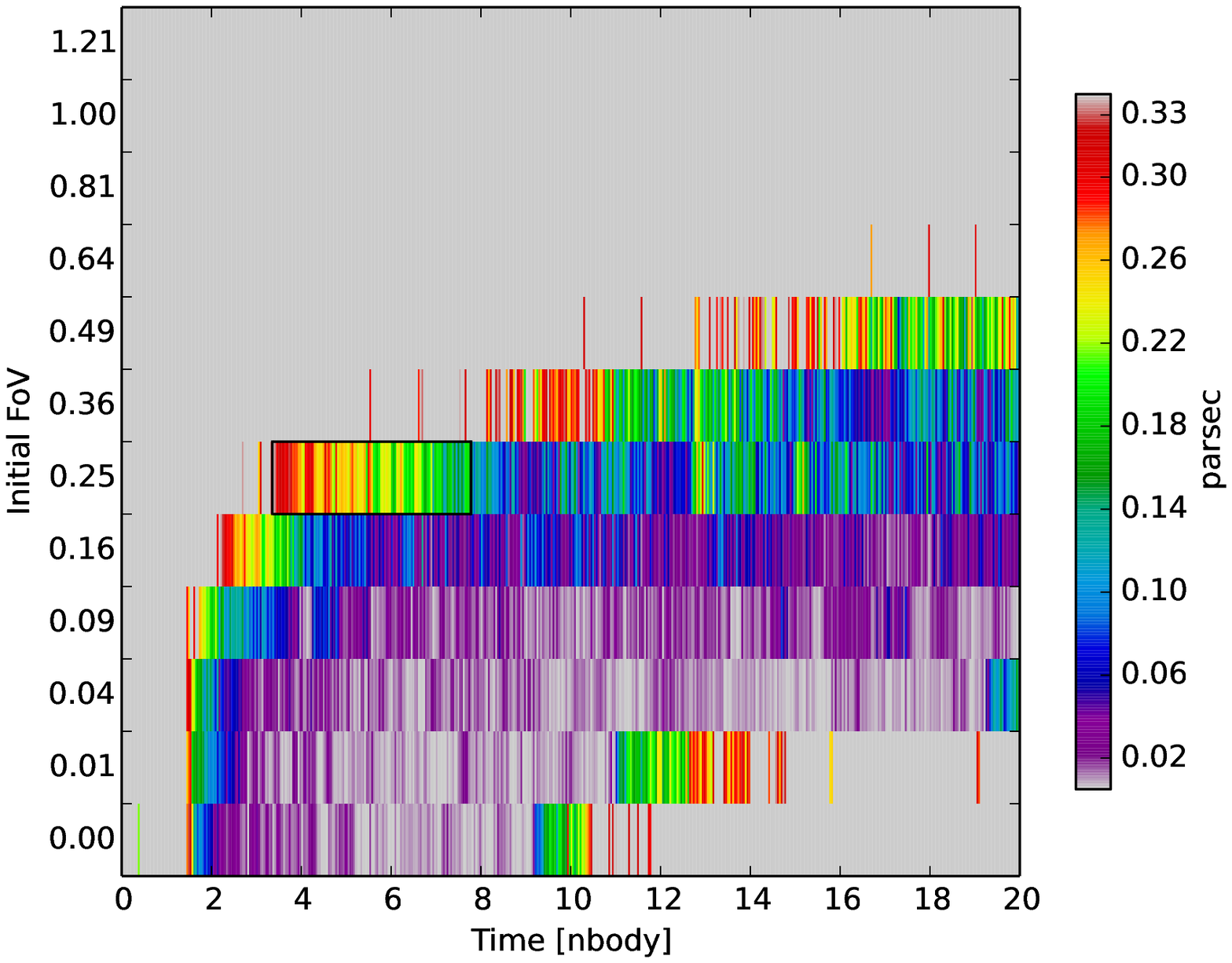}
  }
  \subfloat[Slope of the Density Distribution]{%
    \label{fig:cylinder_density_no_black_hole}
    \includegraphics[width=0.48\textwidth]{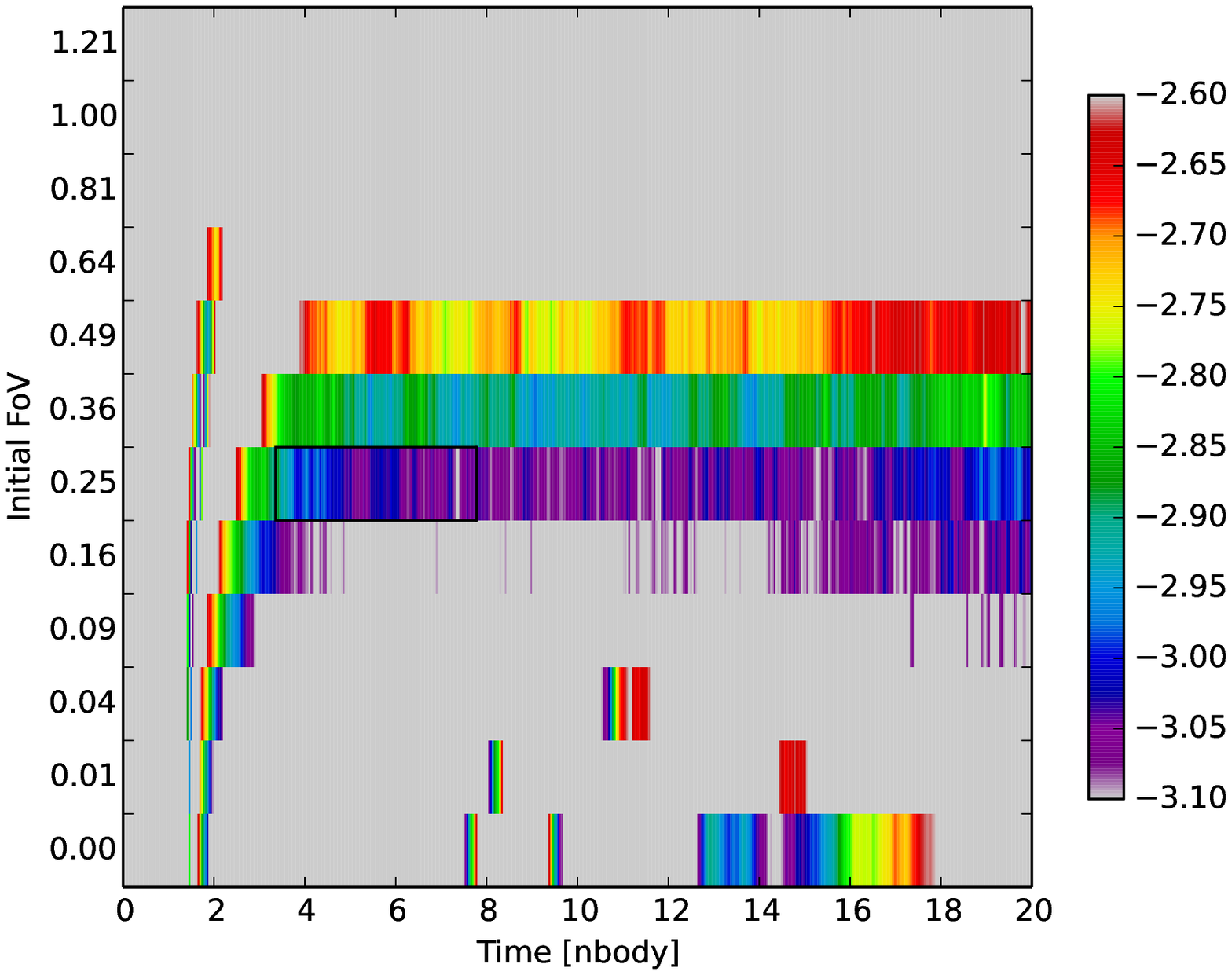}
  }\qquad
  \subfloat[Instantaneous Age]{%
    \label{fig:cylinder_age_no_black_hole}
    \includegraphics[width=0.48\textwidth]{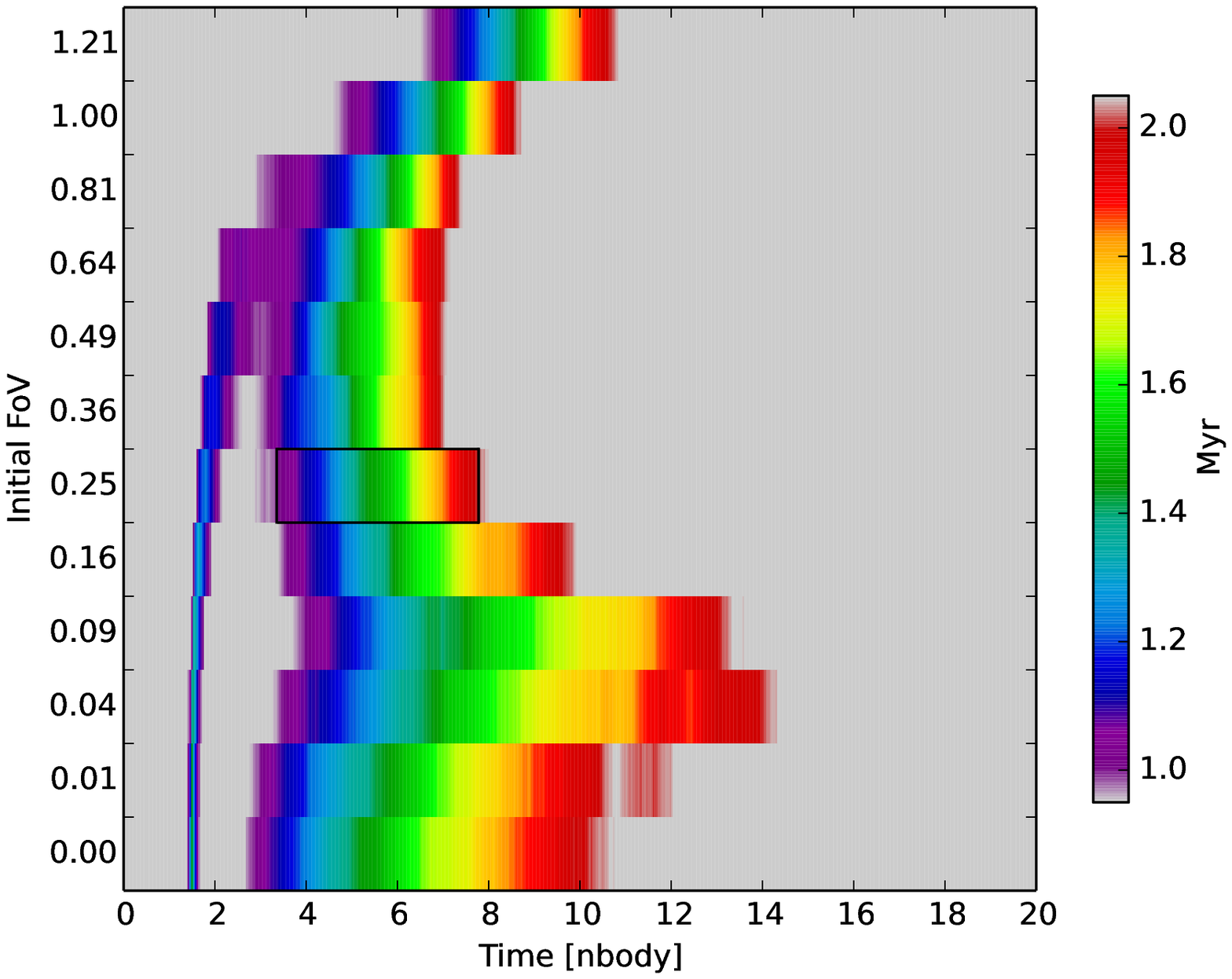}
  }
  \subfloat[Evolved \fov]{%
    \label{fig:cylinder_fov_no_black_hole}
    \includegraphics[width=0.48\textwidth]{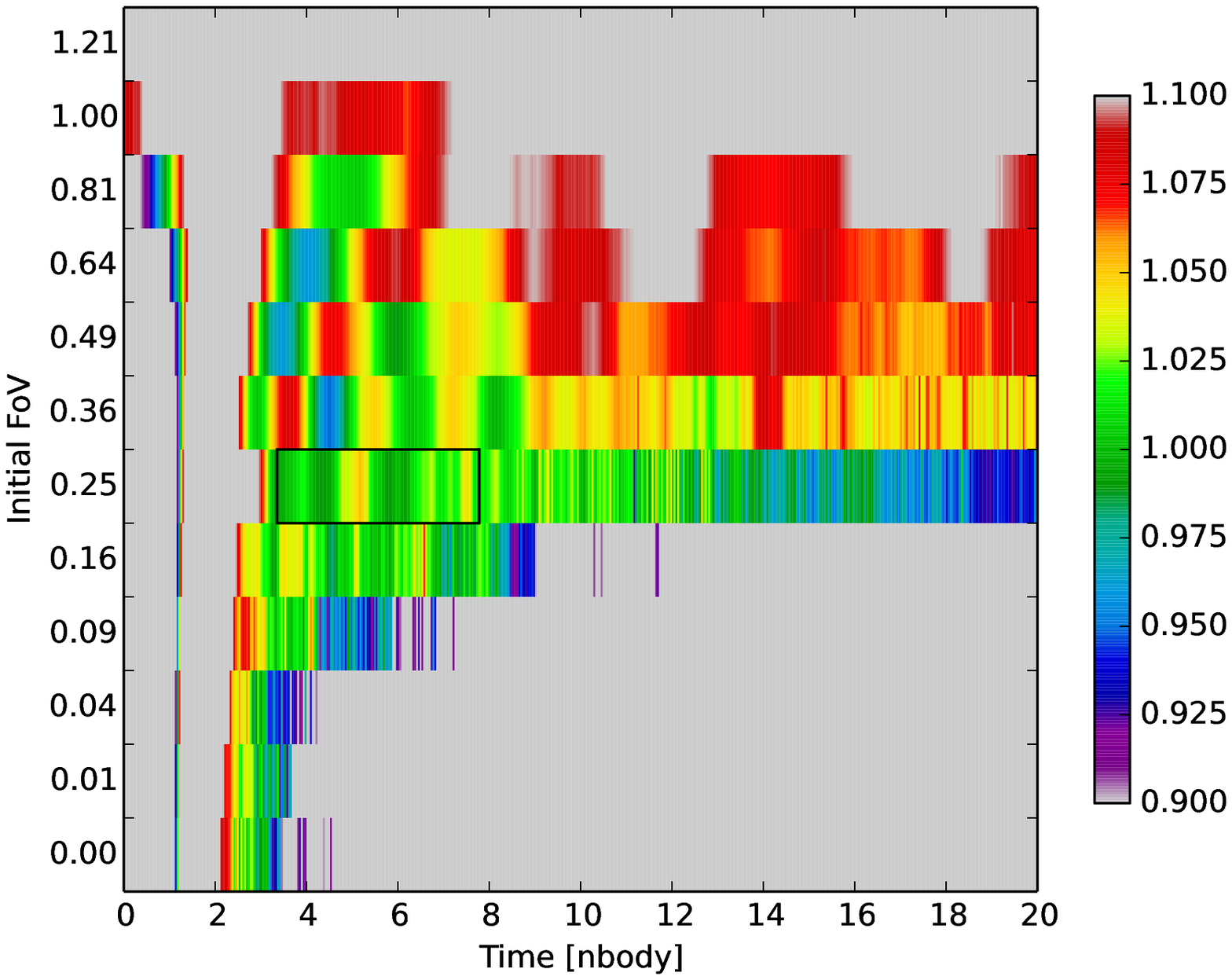}
  }\qquad
  \subfloat[Result (core radius)]{%
    \label{fig:cylinder_result_no_black_hole}
    \includegraphics[width=0.48\textwidth]{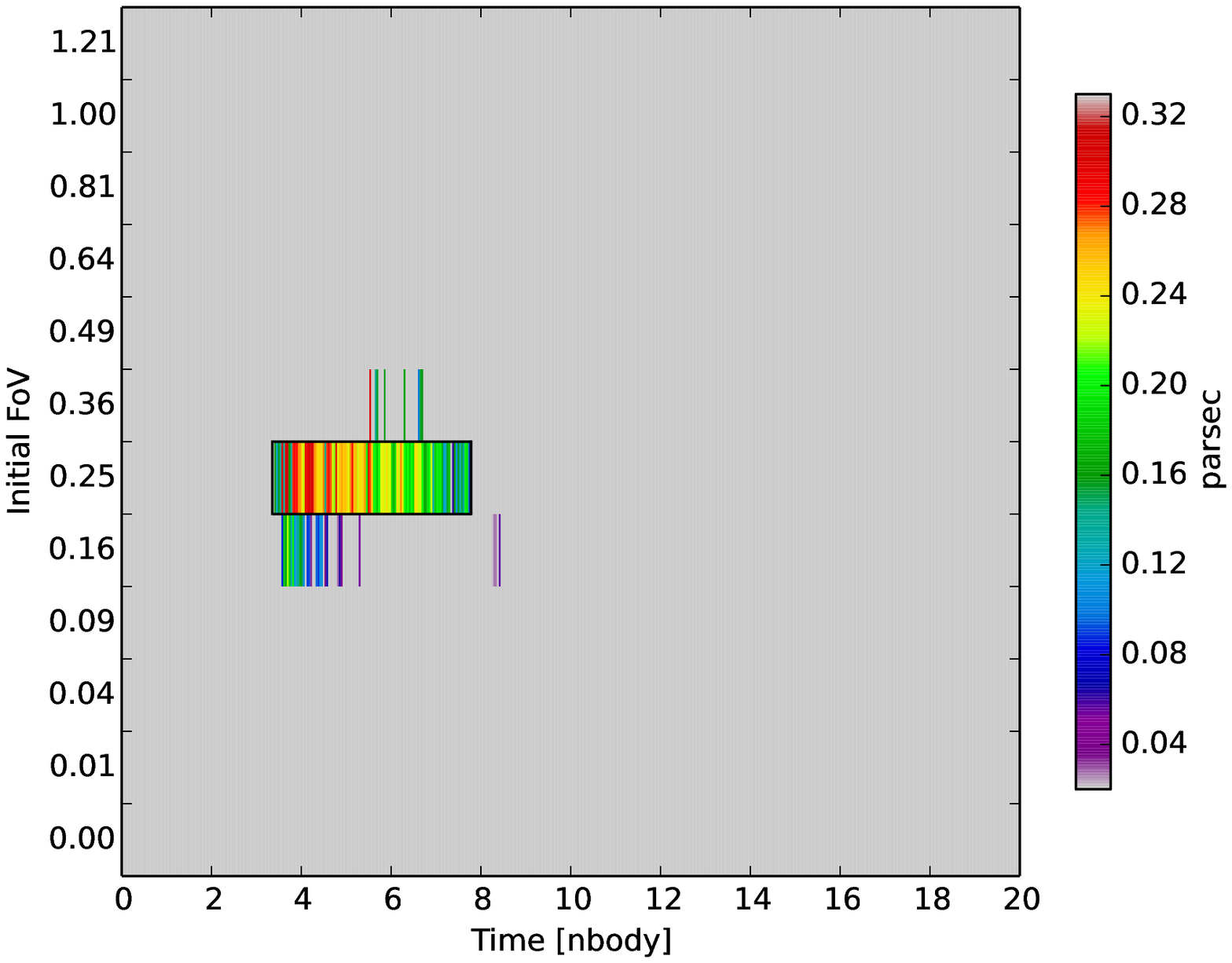}
  }
  \centering
  \caption{\small{
	  Several measures of simulated systems with only the values matching the observational limits of R136 plotted in color.
	  (a) The core radii within observational limits, (b) the slope of the density distribution within observational limits, (c) the age for each snapshot within observational limits, (d) the \fov at each snapshot, and (e) the core radii for snapshots which satisfy all the observational limits. 
          The regions in gray do not produce an accepted value.
          See the text for more information on the limits.
          These systems all began as a homogeneous sphere, with a Salpeter mass function, and without a black hole.
           } }
  \label{fig:cylinder_select_observables_no_black_hole}
\end{figure*}

We define the central region of R136 for our purposes as the volume interior to its virial radius, or $\approx 2.9$~pc (Figure 4 of \citet{Henault2012} presents a nice image of the region with markings for several radii).

We ran simulations (without a black hole, with a Salpeter mass function, and particles initially distributed in a homogeneous sphere) for 20 N-body times.
We show the relevant data in Figure~\ref{fig:cylinder_select_observables_no_black_hole}.
The observables shown in Figure~\ref{fig:cylinder_select_observables_no_black_hole} are not for all bound particles but rather for all particles within a cylinder of radius 2.9~pc from the center of the system; this is done to mimic a projection on to the sky as would be found in the observations.
To reduce noise, we plot the average of every two snapshots thereby reducing the number of data points for each initial \fov from 1000 to 500.

In the top-left panel of the figure, we plot the core radius from our simulations with the color coding, in parsecs, representing the ranges outlined above and values not falling between these ranges are plotted in gray.
In the top-right panel, we plot the slope of the density distribution.
We expect anything within the range of $-2.6$ to $-3.1$ to be consistent with the observed value of $-2.85$ \citep{Harfst2010}.
The middle left panel is a plot of the instantaneous dynamical age of the system with values outside of the measured 1 to 2~Myr plotted in gray.
While we start each simulation with a set \fov it quickly evolves, we have plotted, in the middle right panel, the \fov as it evolves in time.
Since R136 is currently expected to be in virial equilibrium we plot in color the snapshots which have a \fov of $1\pm0.1$.
And finally, in the lowest panel, we show the core radius for only the systems which have a valid measurement for all of the above observables (i.e. core radius, slope of the density distribution, dynamical age, and virial temperature).

We find that within the observational constraints listed above our simulations limit the initial \fov to a likely value between 0.16 and 0.25, with a most likely value of 0.25.
There is also a valid solution at 0.36 but it ranges over a much shorter time and it is not continuous, for this reason we do not find this solution to be as probable.
The continuous solution at 0.16 lasts for nearly $0.2$~Myr whereas the solution at 0.25 lasts for $\approx 1$~Myr so we consider the 0.25 case to be the most likely.

Using the initial \fov of 0.25, we note that in our simulations the core radius is most likely found around 0.2~pc, ranging from about 0.1 to 0.33.
We find very few solutions that allow for the small core radius of the order found in \citet{Hunter1995} or \citet{Andersen2009}.
The range of the simulated \fov is close to 1.0, with the deviations from unity unlikely to be detectable in observations.

We note that these results are based on isolated systems with some simplifications, such as instantaneous star formation and ignoring primordial binaries.
And while R136 is likely to have formed with more complicated initial conditions and is not in isolation, these results provide only a first-order estimate for the initial \fov of R136.
Moreover, this method may be useful when applied to other young clusters which could aid in determining their initial virial temperatures.
We hope this example case has also demonstrated the significance the initial virial temperature has on the evolution of a system.

\subsection{Other Young Clusters}
Using the same analysis techniques we used for R136 we analyzed 15 other extragalactic young clusters.
A list of young clusters within and outside the Local Group can be found in Tables 3 and 4, respectively, of \citet{SPZ2010}.
We required each cluster to have a reported core radius as well as an ``Age$/t_{dyn}$" (the last column in the tables) of less than 20.
Age$/t_{dyn}$ is the inferred age of the cluster divided by dynamical time-scale, or in other words the number of times the typical star has crossed the system [see \citet{Gieles2011} on the usefulness of this measurement].
The clusters which we analyzed are: 3cl-a and a1 in M51; B015D, B040, B257D, B448, and Vdb0 in M31; NGC 1711, NGC 1847, NGC 2004, NGC 2100, NGC 2157, NGC 2164, and NGC 2214 in the LMC; and NGC 330 in the Small Magellanic Cloud.

As these systems are not as well studied as R136 we only used the age and core radius as constraining parameters, but otherwise the analysis remained the same as was performed above for R136.
In 11 of the 15 cases the initial \fov can be fitted well by a value of 0.36 or 0.49 ($Q\approx 0.18$ or $0.25$).
In one case, 3cl-a, the initial \fov is large with a value between 0.64 and 0.81 ($Q\approx 0.32$ or $0.40$), this was the only case with a likely initial \fov greater than 0.49.
In the remaining three cases --- B015D ,B040, and B448 --- the initial \fov was lower than the typical value.
B015D and B448 were best fitted by an initial \fov by 0.04 and 0.09, whereas B040 was best fitted by 0.16 or 0.25.

We find the most typical value (i.e. the mode) for the best-fitting initial \fov in all of the clusters we tested (including R136) to be 0.36 and 0.49 ($Q\approx 0.18$ and $0.25$), collectively these two values fit nearly 70 per cent of the clusters tested.  
A probability-weighted average of the distribution of the initial \fov for these clusters yields a value of 0.30 ($Q= 0.15$).

There may be observational evidence for clusters forming with a rather low initial \fov as we have found here; for example, \citet{Andre2002} studied $\rho$ Ophiuchi and found evidence of collapse.
Additionally, \citet{Walsh2004} found subsonic motion of star-forming cores in NGC 1333 implying subvirial velocities, and \citet{Peretto2006} found signs of global collapse in two massive cluster-forming clumps, namely NGC 2264-C and NGC 2264-D.
Moreover, \citet{Proszkow2009} found that subvirial initial conditions were required in their model in order to explain the kinematic observations of the Orion Nebula Cluster.

\section{Conclusion}
While we suspect that the use of ``cold'' initial conditions is done too often for computational convenience, and with little consideration to physical reasoning, we do not, and cannot, claim that using any particular subvirial temperature is incorrect or less physically consistent since the distribution of the initial virial temperature is unknown.  
We simply aim to demonstrate that the choice of virial temperature is important to consider when formulating initial conditions as this choice has a profound impact on evolution of the resulting cluster.

We also stress the importance of performing analysis only on relevant particles in a simulation, in our case usually the bound particles.
We show an example of the error that can result by analyzing all particles and not only the bound particles in Figure~\ref{fig:all_vs_bound_lagrangian}.
Furthermore, we found that the improper use of unbound particles in the conversion from N-body to physical units lead to unphysical results.

We examined the effect the initial \fov has on the number of particles lost in cases with equal mass particles as well as with a mass function in Figure~\ref{fig:bound_FOV}.  
In the same figure, we find that the addition of a black hole to a cluster has the effect of reducing the number of bound particles after 10 N-body times, as compared to the same system without a black hole, since the black hole acts like a strong scatter.
Additionally, we note an uptick in the number of bound particles for cold systems.
We speculate that this effect is due to the particles initially having no radial motion and so passing through the core on a nearly free-fall trajectory causing them to spend the least amount of time in the very high density core during the collapse.
We then discussed how the mass segregation is dependent on the \fovmath, not only in degree but also in what we called the mass segregation signature (essentially the difference in degree of mass segregation between different mass ranges).

Next we considered the strong influence the choice of initial \fov has on the time-scales (Figure~\ref{fig:timeScales}) and the radius of a system (Figure~\ref{fig:depth}).
In doing so, we find that the minimum radius, $R_{min}$, of a system in violent collapse has a strong dependency on the virial temperature, $Q$, as well as the number of particles, $N$.
We find that $R_{min}\approx Q + N^{(-1/3)}$.
Figure~\ref{fig:timeScales} also provides an estimate to the extra computational expense to reach a mass segregated cluster in a steady state for different initial \fovmath.

After plotting observable quantities, i.e. the core radius, the slope of the density function, and the mass segregation ratio as a function of time and initial \fov in Figures~\ref{fig:no_black_hole_observables}~and~\ref{fig:black_hole_observables}, we discuss the impact of the inclusion of a black hole as the system evolves.
We finally compare our simulated system (particles initially distributed in a homogeneous sphere, with a Salpeter mass function, and without a black hole) to the young cluster R136.
In doing so, we find that given R136's age estimate, the observed current \fovmath, as well as the observed slope of the density distribution, and the many observational constraints on the core radius, R136 would most likely have had an initial \fov of 0.25 ($Q\approx 0.13$).
We repeated the same analysis on 15 other young clusters for which we found 0.36 and 0.49 ($Q= 0.18$ and $\approx 0.25$, respectively) to be the most likely initial \fov in nearly 70 per cent of all 16 young clusters (including R136) and a probability-weighted mean of the distribution of initial \fov to be 0.30 ($Q= 0.15$). 
While these results are robust, we do note that these values are based on an idealized system.

Finally, we hope that this work has convinced the reader of the importance of the initial virial temperature used in simulations.
Whether used as the initial velocities of particles or of merging galaxies, the effect of the virial temperature can be profound and as such should be carefully chosen.

\section*{acknowledgments}
It is a pleasure to thank Steve McMillan for helpful conversations and the anonymous referee for the thoughtful comments.
This work was supported by the Netherlands Research Council (NWO grant numbers 612.071.305 [LGM], 614.061.608 [AMUSE], and 639.073.803 [VICI]) and by the Netherlands Research School for Astronomy (NOVA).

\footnotesize{
  \bibliographystyle{mn2e}
  \bibliography{apj-jour,sub-virial}
}
\end{document}